\documentclass[prd,amssymb,amsmath,floatfix]{revtex4}
\pdfoutput=1 
\usepackage{graphicx}
\usepackage{natbib}

\begin{document}

\title{Life's Chirality From Prebiotic Environments}

\author{Marcelo Gleiser\footnote{Plenary talk delivered at the S\~ao Paulo Advanced School of Astrobiology, S\~ao Paulo, December 2011.}}

\affiliation{Department of Physics \& Astronomy,
Dartmouth College, Hanover, NH 03755, USA
gleiser@dartmouth.edu}

\author{Sara Imari Walker}

\affiliation{NASA Astrobiology Institute, USA}
\affiliation{BEYOND: Center for Fundamental Concepts in Science, Arizona State University, Tempe, AZ 85287 }

\begin{abstract}
A key open question in the study of life is the origin of biomolecular homochirality: almost every life-form on Earth has exclusively levorotary amino acids and dextrorotary sugars. Will the same handedness be preferred if life is found elsewhere? We review some of the pertinent literature and discuss recent results suggesting that life's homochirality resulted from sequential chiral symmetry breaking triggered by environmental events. In one scenario, autocatalytic prebiotic reactions undergo stochastic fluctuations due to environmental disturbances. In another,  chiral-selective polymerization reaction rates influenced by environmental effects lead to substantial chiral excess even in the absence of autocatalysis. Applying these arguments to other potentially life-bearing platforms has implications to the search for extraterrestrial life: we predict that a statistically representative sampling of extraterrestrial stereochemistry will be racemic (chirally neutral) on average. 
\end{abstract}

\maketitle


\section{Introduction}	

During the past few decades the search for life elsewhere in the universe has risen to the forefront of scientific research \cite{Gleiser_Rev}. This is motivated by recent discoveries of exoplanets, including the discoveries of super-Earths \cite{Marcy} and Earth-like planets \cite{Kepler}, opening up the possibility of a potentially large number of habitable planetary platforms beyond Earth. In addition, carbon isotopic evidence indicating that life existed on Earth at least as early as $3.5$ billion years ago (Bya) \cite{VanZuilen02,Schopf93}, and the discoveries of extremophilic life forms on Earth \cite{SRS}, suggest that life can survive and even thrive under harsher conditions than previously imagined. In light of such evidence, it is reasonable to conjecture that at least simple, unicellular life may be more common in the universe than anticipated. When attempting to answer the question of how widespread life is, it is pertinent to examine the only example of abiogenesis known to date: the origin of life on Earth.
 
One of the most distinctive features of life--the existence of a specific and seemingly universal chiral signature--also presents one of the longest standing mysteries in studies of abiogenesis. It is well--known that chiral selectivity plays a key role in the biochemistry of living systems: nearly all life on Earth contains exclusively dextrorotary sugars and levorotary amino acids. Quite possibly, the development of homochirality was a critical step in the emergence of life. Although there are numerous models for the onset of homochirality presented in the literature, none is conclusive: the details of chirobiogenesis remain unknown. 

We propose that the environment of early Earth played a crucial role in chirobiogenesis. Environmental effects, when strong enough, can destroy any memory of a prior chiral bias, whatever its origin.  They can also drive a specific chiral choice. Life's chirality is interwoven with early-Earth's environmental history; specifically, with how the environment influenced the prebiotic soup that led to first life. The same should be true of life anywhere in the cosmos. As remarked in Ref. \cite{Gleiser_Rev}, {\it the history of life in a planet mirrors the planet's life history.} Here, it will be argued that the same is true for life's chirality.

\section{Life in the Universe}

The rapidity with which life first appeared on Earth is often cited as evidence that life may be common in the universe. Observational constraints on the timescale for the origin of life on Earth suggest that Earth-like planets older than about $1$ Gyr support their own abiogeneses with a probability $>13 \%$ at the $95 \%$ confidence level \cite{LD}. Combining this estimate with results demonstrating that Earth--like planets in habitable zones may be a common byproduct of star formation \cite{Kasting,Line}, the search for extraterrestrial life may yield promising results in the near future. However, until extraterrestrial life is discovered, the existence of life on Earth is our only insight into abiogenesis, presenting important clues on the likelihood of life elsewhere in the universe.

Given early Earth's tumultuous environment, it is remarkable that paleontological evidence suggests life may have been thriving as early as $3.5$ Bya \cite{VanZuilen02,Schopf}. Perhaps even more surprising are results suggesting that any early life may have been killed off as late as $3.8$ Bya \cite{MS,Sleep}. These findings indicate that the origin and early diversification of life occurred within a window as short as $300$ million years. Additional constraints stemming from the half--life of prebiotic compounds and the recycling of oceans through hydrothermal vents set an even shorter timescale for abiogenesis, estimated to be as brief as $5$ million years \cite{Lazcano}.

The short timescales constraining the origin of life on Earth shed little light on the environmental conditions required for abiogenesis. The question of where life began remains open, with few details known about where the first prebiotic ingredients may have been synthesized. Potential sites for the origin of life include submarine vents \cite{Corliss81}, primordial beaches \cite{Bywater}, and shallow pools and lagoons \cite{RM}. An even more debated aspect of the puzzle is whether the ingredients for life originated on Earth or were delivered to Earth from outerspace \cite{Chyba}. Evidence supporting the latter hypothesis includes the ubiquity of organic chemical ingredients in the interstellar medium \cite{Charnley} and in the solar system in carbonaceous chondrites \cite{Cronin}. However, organic compounds have also been shown to be readily synthesized under conditions likely to be prevalant on the prebiotic Earth \cite{StanM,RWFM,WHM}.

Of equal ambiguity is the mechanism of abiogenesis. Although the uniformity of life on Earth suggests that all extant organisms descended from a last universal common ancestor (LUCA), we know almost nothing about the abiotic ingredients and prebiotic chemistries present on the primitive Earth from which the LUCA evolved  \cite{O}. Potential mechanisms range from ``metabolism-first'' models, such as the iron-sulfide world hypothesis of W\"achtersh\"auser \cite{Wach} and ``membrane--first'' lipid-world scenarios as investigated by Deamer and coworkers \cite{MD,MBD}, to the ``peptide-first'' models proposed by Fox \cite{Fox,Fox2} and others \cite{Fishkis,GW2}, and the popular ``genetics-first'' hypotheses such as the RNA \cite{Gilbert} and pre-RNA \cite{Orgel2000} world scenarios. In considering any of these models, one must investigate how the characteristic properties of life might have arisen--including the emergence of homochirality.

With such a short window for the origin of life, it is likely that the primordial Earth experienced multiple abiogeneses. As has been pointed out by Davies and Lineweaver \cite{DL}, if large impacts had frustrated abiogenesis, then as the frequency of impacts abated at the end of heavy bombardment, there would have been brief quiescent periods when life may have emerged only to be annihilated by the next large impact. Extending this to studies of potential abiogenic mechanisms, we must therefore be mindful that in prebiotic Earth reactor pools were submitted to environmental disturbances ranging from mild (e.g. tides, evaporating lagoons) to severe (e.g. volcanic eruptions, meteoritic impacts). Both kinds of disturbances must have affected the evolution of chirality in early Earth \cite{GTW,BM,Hochberg}. For the remainder of this work, we will discuss how such a scenario might have played out.

\section{Deciphering the Origin of Life's Chirality}

A much-debated question is whether the observed homochirality of biomolecules is a prerequisite for life's emergence or if it developed as its consequence \cite{Cohen,Bonner}. Adding to the mystery, prebiotically relevant laboratory syntheses yield racemic mixtures \cite{Dunitz}. This is especially surprising given that statistical fluctuations of reactants will invariably bias one enantiomer over the other \cite{Blackmond}: every synthesis is {\it ab initio} asymmetric. It is thus clear that this asymmetry is erased as the reactions unfold. Therefore, a chiral selection process must have occurred at some stage in the origin or early evolution of life \cite{Bada}. 

A common viewpoint is that chiral selection occurred at the molecular level \cite{Lahav}, and that the resultant complexity of molecular species led to the eventual emergence of life. This view is based on the argument that life could neither exist nor originate without biomolecular asymmetry \cite{Bonner}, and is supported by experiments demonstrating that specific conformations of structural entities such as $\alpha$-helices and $\beta$-sheets can only form from enantiomerically-pure building blocks \cite{Cline,Fitz} (for an alternative viewpoint see Ref. \cite{Nielsen}). Taking this bottom-up approach, we therefore assume that the prebiotic conditions necessary for the subsequent development of complex biomolecules had to be chiral.

\subsection{Modeling Prebiotic Homochirality}

Although it was Louis Pasteur \cite{Pasteur}, inspired by previous work by Biot, who in the late 1840s was the first to recognize  that many biomolecules display mirror asymmetry, it was not until the pioneering work of Frank \cite{Frank}, over one--hundred years later, that the first breakthrough in understanding the origin of this asymmetry was presented. In this influential work, Frank identified autocatalysis and some form of mutual antagonism as necessary ingredients for obtaining biomolecular homochirality from prebiotic precursors. In the ensuing decades, many models exhibiting such features have been proposed, each providing its own description of chiral symmetry breaking.

The various models presented in the literature range from investigations of simple modifications to Frank's original model \cite{KN83,Hochberg07}, to more recent studies describing the onset of homochirality in crystallization \cite{SH05,Viedma}, and chiral selection during polymerization \cite{Sandars,SH} (see Ref.~ \cite{Plasson07} for a detailed discussion). Among these ``Frank'' models, one of the better known is that of Sandars \cite{Sandars}, which provides a basis for understanding chiral symmetry breaking in a RNA world \cite{Nilsson}. This model succeeds because it includes both of the necessary features of antagonism and autocatalysis as originally proposed by Frank, where the mutual antagonism is provided by enantiomeric cross-inhibition as is observed in template--directed polycondensation of polynucleotides \cite{Joyce}. (These terms will be clarified below.)

As various authors have pointed out \cite{GW2,Plasson07,BLL}, when addressing the validity of the Sandars model one must consider that the autocatalysis necessary for chiral symmetry breaking in such systems is presently only observed for a few non-biological molecules \cite{Blackmond,Soai}, and would be trying to achieve with even very simple organic molecules \cite{Joyce2}. Despite this shortcoming, the RNA world hypothesis is still deemed viable by some authors \cite{Monnard}. In addition, the Sandars model provides an elegant, and relatively simple, model of chiral symmetry breaking while sharing general features in common with other (usually more complicated) models, and as such it has been extensively studied in the literature \cite{BM,G}. As we know little about the compositions of the primitive atmospheres and seas \cite{Lazcano} and even less about prebiotic chemistry \cite{Orgel}, it is pertinent to study general features as opposed to details of specific models. We thus begin by investigating the Sandars model as the basis of our study with the reasonable expectation that the results should be qualitatively similar for other models. We will then briefly describe a recently developed polymerization model that obtains a substantial amount of chiral bias without any specific external source: only the reaction rates display chiral dependence which may be induced environmentally \cite{GNW}.

\subsection{Prebiotic Homochirality as a Critical Phenomenon}

While the details of models describing the onset of prebiotic homochirality mentioned above differ, the qualitative features are the same; chiral symmetry breaking occurs due to the introduction of instabilities to the symmetric (racemic) state that lead to spontaneous symmetry breaking in physical systems \cite{Plasson07,G}. In other words, the spatiotemporal dynamics of the reaction network is equivalent to a two-phase system undergoing a symmetry-breaking phase transition, where the order parameter is the net chiral asymmetry, ${\cal A}$. If we define ${\cal L}$ and ${\cal D}$ as the sums of all left and right--handed chiral subunits, respectively, then the net chirality may be defined as
\begin{eqnarray} \label{A}
{\cal A} = \frac{{\cal L} - {\cal D}}{{\cal L} + {\cal D}}.
\end{eqnarray}
Note that the net chirality is symmetric ${\cal A}_0 = 0$ in the racemic state, and asymmetric ${\cal A}_{+,-} \neq 0$ in the non-racemic states. The reaction network is a nonlinear dynamical system with behavior controlled by model--dependent parameters, including fidelity of enzymatic reactions \cite{Sandars,BM,WC}, ratios of reaction rates \cite{GW}, stereoselectivity \cite{Plasson04} and total mass \cite{GW2}.

Just as in other areas of physics, environmental interactions can restore the system to the symmetric (racemic) state, even in cases where model parameters are set such that the asymmetric state is stable. In such cases, it is important to consider how temperature, or other environmental effects, might work to restore the stability of the racemic state. Thinking in this direction, it was Salam \cite{Salam} who first suggested that there should be a critical temperature, $T_c$, above which any net chirality is destroyed. One can think in analogy with a ferromagnet: if heated through the Curie point any net magnetization is erased and the system is restored to a symmetric configuration. Here, the net chirality plays the role of the net magnetization. While Salam conceded that calculating $T_c$ would be challenging using the electroweak theory of particle physics (assuming the weak force biases chiral selection \cite{Yamagata,KN85}), a different route was recently taken by Gleiser and Thorarinson \cite{GT}. Coupling the reaction network to an external environment modeled by a stochastic force, they were able to determine the critical point for homochirality in two and three dimensions. We move now to a discussion of their work.

\subsubsection{Modeling Spatiotemporal Polymerization}

Although the work of Gleiser and Thorarinson was based on the Sandars model, from the above discussion and the work of Gleiser and Walker \cite{GW} we expect the results  to be quite general. The reaction-network proposed by Sandars includes the following polymerization reactions:
\begin{eqnarray}\label{rxnnetwork}
L_n + L_1 & \stackrel{2k_S}{\rightarrow} L_{n+1}, \nonumber \\
L_n + D_1 & \stackrel{2k_I}{\rightarrow} L_nD_1, \nonumber \\
L_1 + L_nD_1 & \stackrel{k_S}{\rightarrow} L_{n+1}D_1, \nonumber \\
D_1 + L_nD_1 & \stackrel{k_I}{\rightarrow} D_1L_nD_1, 
\end{eqnarray} 
supplemented by reactions for $D$-polymers by interchanging $L \rightleftharpoons D$ (consistent with biochemical usage, we denote $D$-compounds by the letter ``$D$'' as opposed to the notation set in Sandars' work where such molecules were denoted as ``$R$''). A left-handed polymer $L_n$, made of $n$ left-handed monomers, $L_1$, may grow by adding another left-handed monomer with a rate $k_s$, or be inhibited by adding a right-handed monomer $D_1$ with a rate $k_I$. The latter process is referred to as enantiomeric cross-inhibition: attachment of a monomer with opposite chirality to one end of a growing chain terminates growth on that end of the chain \cite{Joyce} and inhibits any further enzymatic activity. This process is the driving force that causes a net asymmetry to develop in this model \cite{GW}.

In addition, the reaction network includes a substrate, $S$, from which monomers of both chiralities are generated: $S \stackrel{k_C(p C_L + q C_D)}{\longrightarrow} L_1$;  $S \stackrel{k_C (p C_D + q C_L)}{\longrightarrow} D_1$, where $p = \frac{1}{2}(1+f)$ and $q = \frac{1}{2}(1-f)$, with $f$ a measure of the enzymatic fidelity. $C_{L (D)}$ determine the enzymatic enhancement of chirally-pure $L (D)$-handed monomers, and are assumed to depend on the length of the largest polymer in the reactor pool, $N$, such that $C_{L(D)} = L_N(D_N)$ \cite{Sandars}. Other choices are possible, but lead to similar qualitative results \cite{GW}.

Given that the Soai reaction \cite{Soai} - the most well--known illustration of an autocatalytic network leading to chiral purity - features dimers as catalysts \cite{Blackmond}, we focus on the truncated system for $N = 2$. We note that it is possible to make this truncation while maintaining the essential aspects of the dynamics leading to homochiralization \cite{GW}. The reaction network is further simplified by assuming that the rate of change of the substrate, $[S]$, and of the dimers, $[L_2]$ and $[D_2]$, is much slower than that of the monomers, $[L_1]$ and $[D_1]$. These approximations are known as the adiabatic elimination of rapidly adjusting variables \cite{Haken} and have been shown to produce a reliable approximation to the full ($n>2$) Sandars model \cite{GW}. 

It is convenient to introduce the dimensionless symmetric and asymmetric variables, ${\cal S} \equiv X + Y$ and ${\cal A} \equiv X - Y$, where $X \equiv [L_1](2 k_S/ Q_S)^{1/2}$ and $Y \equiv [D_1](2 k_S/ Q_S)^{1/2}$, respectively \cite{BM}. For $k_I/k_S=1$, after a little algebra, the reaction network simplifies to
\begin{eqnarray} \label{eqns}
l_0^{-1} \frac{d{\cal S}}{dt} &=&  1 - {\cal S}^2,   \\ \nonumber
l_0^{-1} \frac{d{\cal A}}{dt}  &=& \frac{2 f {\cal S}{\cal A} }{{\cal S}^2 + {\cal A}^2} -{\cal S}{\cal A}, 
\end{eqnarray}
where $l_0 \equiv (2k_SQ)^{1/2}$ has the dimensions of inverse time. ${\cal S} = 1$ is a fixed point: the system tends quickly toward this value at time--scales of order $l_0^{-1}$. 

Substituting ${\cal S} =1$ into eqs. \ref{eqns}, we obtain an effective potential for $V({\cal A})$,
\begin{eqnarray} 
V({\cal A}) = \frac{{\cal A}^2}{2} - f \ln \left[ {\cal A}^2 + 1\right] ,
\end{eqnarray}
with fixed points ${\cal A} = 0, \pm \sqrt{2f -1}$. Note that for $f < 1/2$ an enantiomeric excess is impossible and the only steady state is the (stable) symmetric state. In the case $f=1$, the potential takes the form of a symmetric double--well, where the two fixed asymmetric steady states are homochiral (${\cal A} =  \pm 1$) and represent the global minima. In this case, the symmetric state is unstable.

The form of this potential introduces the possibility of describing chiral symmetry breaking as a phase transition. This, in fact, suggests that a proper treatment of the problem should include spatial dependence. To introduce spatial dependence to the reaction network, the usual procedure in the phenomenological treatment of phase transitions is implemented with the substitution $d/dt \rightarrow \partial / \partial t - k \nabla^2$, where $k$ is the diffusion constant \cite{BM}. In this coarse-grained approach, the number of molecules per unit volume is large enough so that the concentrations vary smoothly in space and time. Dimensionless time and space variables are then defined as $t_0 = l_0 t $, and $x_0 = x(l_0/k)^{1/2}$, respectively. For diffusion in water, $k = 10^{-9}$m$^2$s$^{-1}$, and nominal values $k_S = 10^{-25}$cm$^3$s$^{-1}$ and $Q = 10^{15}$ cm$^{-3}$s$^{-1}$, we obtain $l_0=\sqrt{2} \times 10^{-5}$s$^{-1}$ corresponding to $t \simeq (7 \times 10^4$s$) t_0$ and $x \simeq (1$cm$)x_0$. 

\begin{figure}
\centerline{\includegraphics[width=0.6\linewidth]{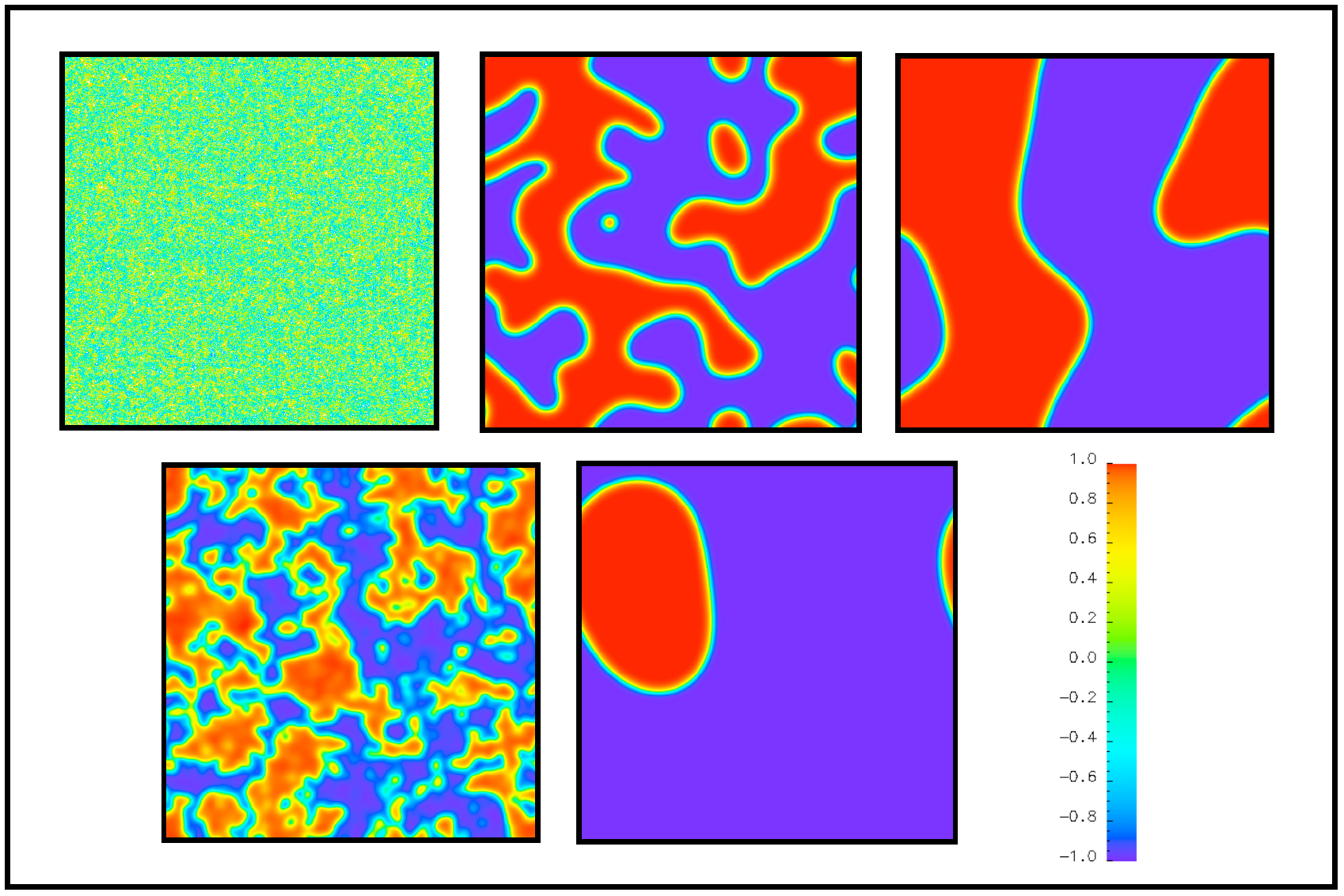}}
\caption{Evolution of $2d$ chiral domains. Red (+1 on the color bar) corresponds to the $L$-phase and blue (-1 on the color bar) corresponds to the $D$-phase. Time runs from left to right and top to bottom. Top left, the near-racemic initial conditions. Top mid and top right, evolution of the two percolating chiral domains separated by a thin domain wall. Bottom left, environmental effects break the stability of the domain wall network. Bottom right, subsequent surface-tension driven evolution leads to a enantiomerically-pure world. } \label{fig:2Dchiral}
\end{figure}

\begin{figure}
\centerline{\includegraphics[width=0.6\linewidth]{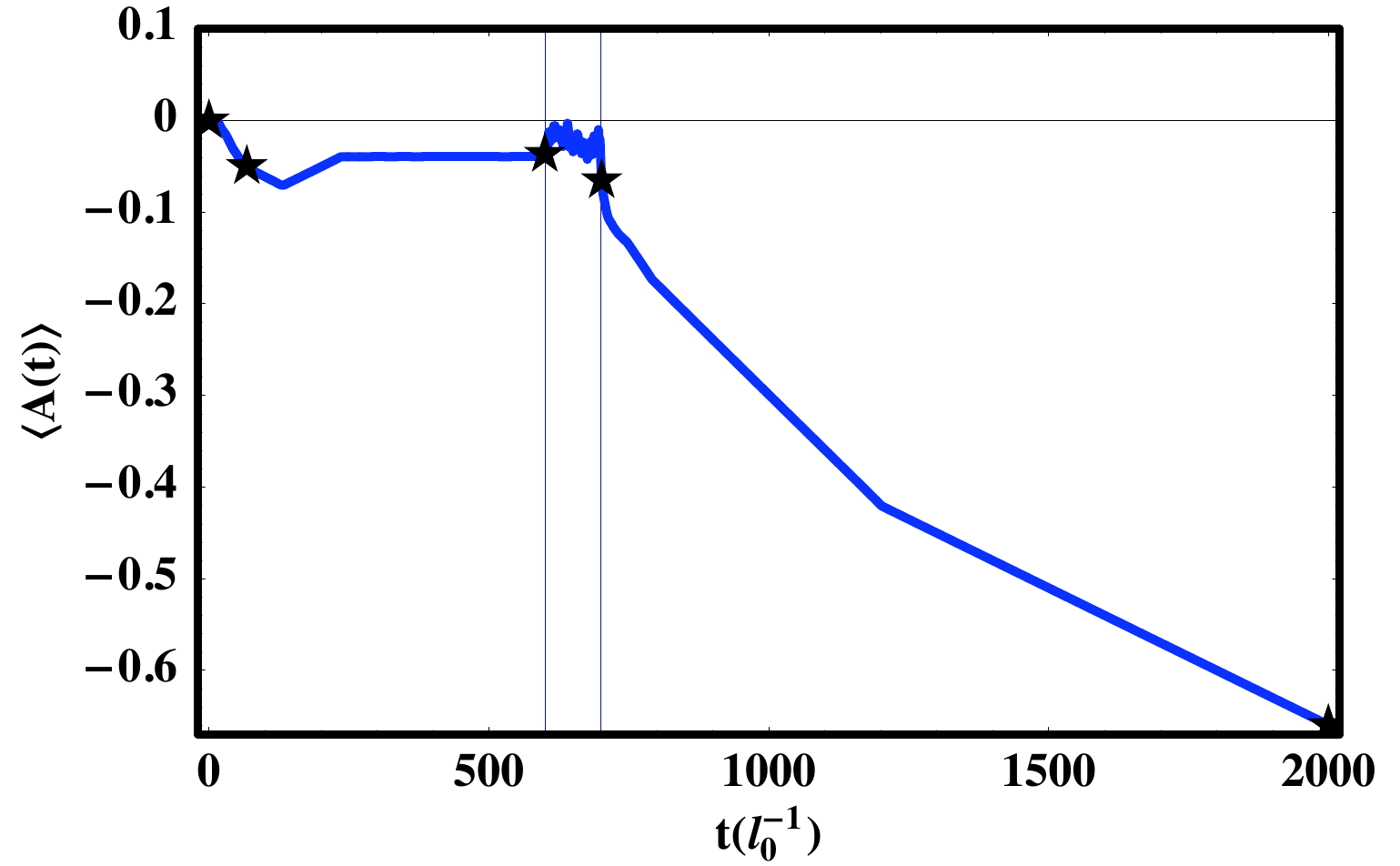}}
\caption{Time evolution of the spatially-averaged net chirality corresponding to the snapshots shown in Figure \ref{fig:2Dchiral}. Stars denote times for snapshots and vertical lines mark the beginning and end of stochastic environmental  influence.} \label{fig:Aave}
\end{figure}

Considering the case where $f=1$, for near-racemic initial conditions ($ |{\cal A}(0,x,y,z)| \leq10^{-4}$), the spatiotemporal evolution leads to the formation of left and right-handed percolating chiral domains separated by domain walls, as is well-known from systems in the Ising universality class (see Fig. \ref{fig:2Dchiral}). Surface tension drives the walls until their average curvature matches approximately the linear dimension of their confining volume. At this point, wall motion becomes quite slow, $d \langle {\cal A}(t) \rangle /dt \rightarrow 0$, where $\langle {\cal A}(t) \rangle$ is the spatially-averaged value of the net chiral asymmetry, and the domains coexist in near dynamical equilibrium in that the net stresses add to zero (see Fig. \ref{fig:2Dchiral}, top right). The time evolution of ${\cal A}(t)$ is shown in Fig \ref{fig:Aave}. For such model systems, and considering only diffusive processes, it has been shown that the presence of a bias from parity-violating weak neutral currents (PV) or most circularly-polarized light (CPL) sources \cite{Lucas} (even in the unlikely situation where they could be sustained unperturbed for hundreds of millions of years), would not lead to chirally-pure prebiotic conditions \cite{G}. 

\subsubsection{Coupling to the Environment: A Critical Point for Homochirality}

Chiral symmetry breaking in the context of this model can be understood in terms of a second-order phase transition, where the critical ``temperature'' is determined by the strength of the coupling between the reaction network and the external environment. The external environment is modeled via a generalized spatiotemporal Langevin equation \cite{GT} by rewriting eqns. \ref{eqns} as,
\begin{eqnarray} \label{spatialeqns}
l_0^{-1}\left( \frac{\partial{\cal S}}{\partial t} - k \nabla^2 {\cal S} \right) &=&  1 - {\cal S}^2 + w(t, \textbf{x}), \nonumber \\ 
l_0^{-1}\left( \frac{\partial{\cal A}}{\partial t} - k \nabla^2 {\cal A} \right) &=& {\cal S}{\cal A} \left( \frac{2f}{{\cal S}^2 + {\cal A}^2} -1 \right)  +  w(t, \textbf{x}), 
\end{eqnarray}
where $l_0 \equiv (2k_SQ)^{1/2}$, and $w(\textbf{x},t)$ is a dimensionless Gaussian white noise with two-point correlation function $\langle w(\textbf{x}',t')w(\textbf{x},t) \rangle = a^2\delta(t'-t) \delta(\textbf{x}' -\textbf{x})$. The parameter $a^2$ is a measure of the environmental influence. For example, in the mean-field models of phase transitions, it is common to write $a^2 = 2 \gamma k_B T$, where $\gamma$ is the viscosity coefficient, $k_B$ is Boltzmann's constant, and $T$ is the temperature. Using the dimensionless space and time variables, $t_0 = l_0 t $, and $x_0 = x(l_0/k)^{1/2}$, introduced above, the noise amplitude scales as $a_0^3 \rightarrow \lambda_0^{-1} (\lambda_0 /k)^{d/2}a^2$, where $d$ is the number of spatial dimensions. A crucial point is that even in the case of perfect fidelity, $f = 1$, where the potential supports stable homochiral steady-states, an enantiomeric excess may not develop if $a$ is above a critical value $a_c$. 

\begin{figure}
\centerline{\includegraphics[width=0.6\linewidth]{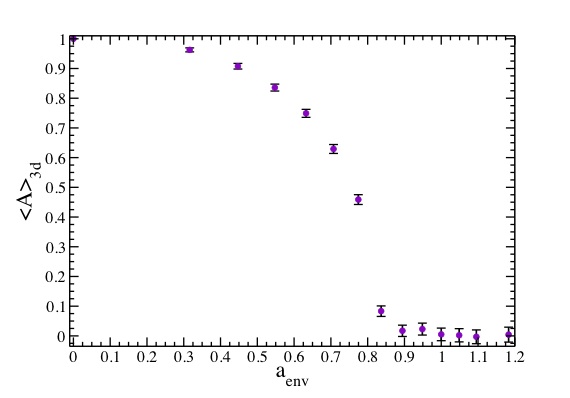}}
\caption{Average enantiomeric excess versus environmental "temperature" in three dimensions. The error bars denote ensemble averages over $20$ runs.} \label{fig:GT}
\end{figure}

As shown in Fig. \ref{fig:GT}, an Ising phase diagram can be constructed showing that $\langle {\cal A} \rangle \rightarrow 0$ for $a > a_c$ and chiral symmetry is restored. The value of $a_c$ has been obtained numerically in two ($a_c^2 = 1.15(k/l_0^2)$cm$^2$s) and three ($a_c^2 = 0.65(k^3/l_0^5)^{1/2}$cm$^3$s) dimensions \cite{GT}. Above $a_c$ the stochastic forcing due to the external environment overwhelms any local excess of $L$ over $D$ within a correlation volume $V_\xi \sim \xi^d$, where $\xi$ is the correlation length: racemization is achieved on large scales and chiral symmetry is restored throughout space. 

In light of these results, it has been shown that within the violent environment of prebiotic Earth, effects from sources such as weak neutral currents (which introduce a small tilt in the potential), even if cumulative, would be negligible: any accumulated excess could be easily wiped out by an external disturbance \cite{GTW,GT}. The history of life on Earth and on any other planetary platform is inextricably enmeshed with its early environmental history.

\section{Punctuated Chirality \cite{GTW}}

The results of the previous section indicate that the environment of early Earth, or other potential prebiotic extraterrestrial environments, must have played a crucial role in chirobiogenesis. The chirality of the prebiotic soup might have been reset multiple times by significant environmental events such as active volcanism and meteoritic bombardment. Under this view, the history of prebiotic chirality is interwoven with the Earth's environmental history through a mechanism we call {\it punctuated chirality} \cite{GTW}: life's homochirality resulted from sequential chiral symmetry breaking triggered by environmental events. 

Punctuated chirality is an extension of the the punctuated equilibrium hypothesis of Eldredge and Gould \cite{Eldredge} to prebiotic times. The theory of punctuated equilibrium describes evolutionary processes whereby speciation occurs through alternating periods of stasis and intense activity prompted by external influences: the punctuation is the geological moment when species arise which may be slow by human standards but is certainly abrupt by planetary standards as evidenced by the fossil record. If phyletic gradualism (traditional Darwinian evolution) is like pushing a ball up an inclined plane, then punctuated equilibrium is like climbing a staircase \cite{Gould}.

It is commonly accepted that molecules undergo selective processes that lead to evolutionary adaptions (see for example Refs. \cite{Kimura}, \cite{Trevors}). It is therefore natural to extend the punctuated equilibrium hypothesis to the prebiotic realm. In this context, the concept of punctuated equilibrium is borrowed with some freedom: the network of chemical reactions described in prebiological systems is a non-equilibrium open system capable of exchanging energy with the environment. The periods of stasis that develop correspond to steady-states in that even though environmental influences may be negligible, chemical reactions are always occurring so as to keep the average concentrations of reactants at a near-constant value. 

\begin{figure}
\centerline{\includegraphics[width=0.6\linewidth]{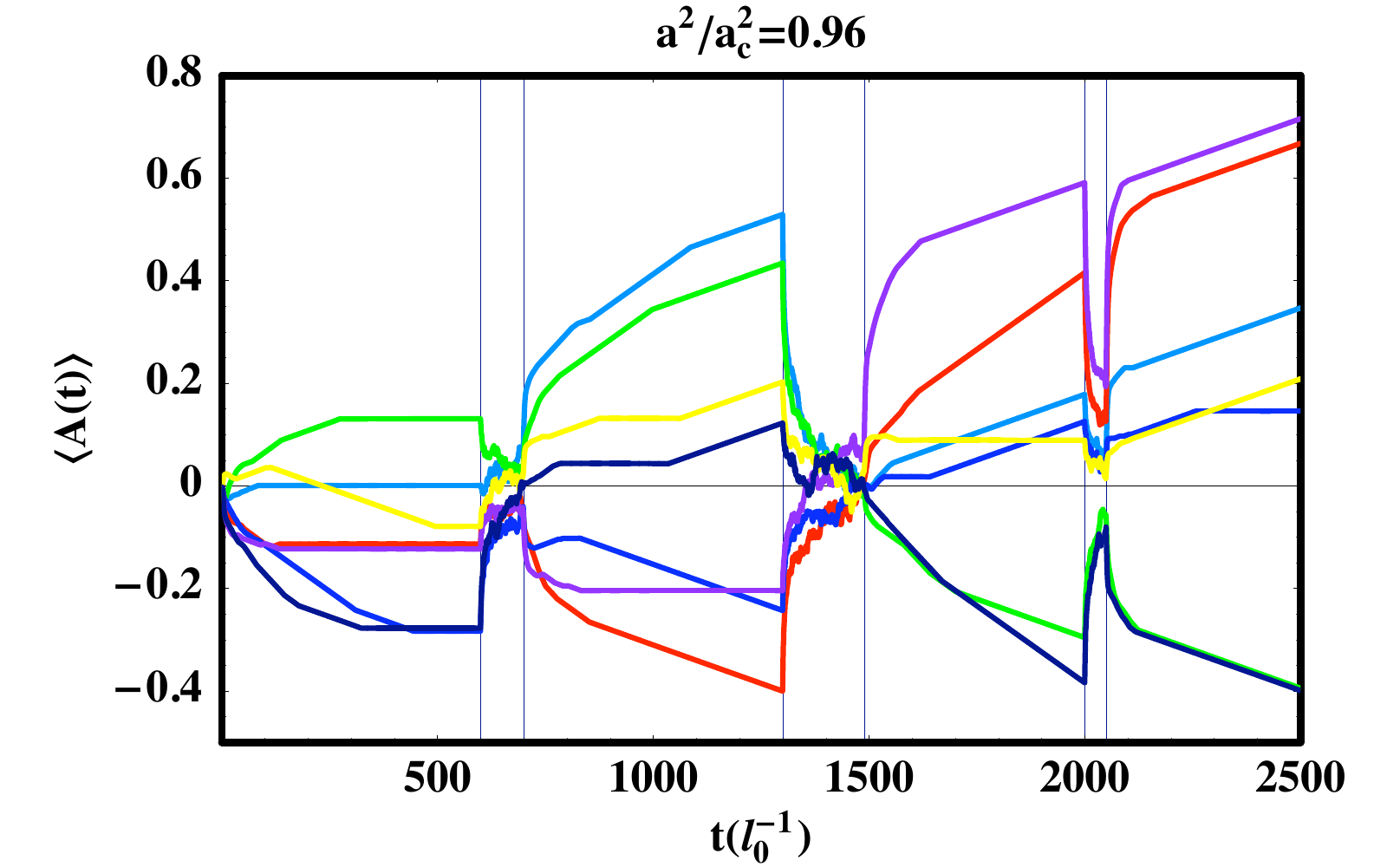}}
\caption{Punctuated Chirality. Impact of environmental effects of varying duration and fixed magnitude ($a^2/a_c^2 = 0.96$) on the evolution of prebiotic chirality in $2d$. Short events (last from left), which have little to no effect, should be contrasted with longer ones, which can drive the chirality towards purity and/or reverse its trend. (See, {\em e.g.} the green line.)} \label{fig:3event}
\end{figure}

As an example of punctuated chirality in a prebiotic scenario, one can consider how repeated environmental interactions influence the evolution of chirality in the context of the model presented in the previous section. In Fig. \ref{fig:3event}, we show several $2d$ runs where the environmental effects vary in duration, while their magnitude was set at $a^2/a_c^2 = 0.96$, so that the magnitude of the disturbance is just below the critical value found by Gleiser and Thorarinson \cite{GT}. Each colored line represents a prebiotic scenario, with the same environmental disturbances of different duration occurring in sequence. 

In order to investigate the impact of environmental effects on chiral selectivity, the scenarios reflect situations where there is no chiral selection, that is, where the two phases coexist in dynamical equilibrium (mathematically, when $d \langle {\cal A}(t) \rangle /dt \rightarrow 0$ for ${\cal A}(t) \neq \pm 1$; chemically, in a steady state). We observe that long disturbances can drive the net chirality towards purity ($ \langle {\cal A}(t) \rangle \rightarrow  \pm 1$ for large $t$). Furthermore, note that subsequent events may erase any previous chiral bias, favoring the opposite handedness. In other words, environmental effects of sufficient intensity and duration can reset the chiral bias. This is true even if the system evolves toward homochirality prior to any environmental event.

\begin{figure}
\centerline{\includegraphics[width=0.6\linewidth]{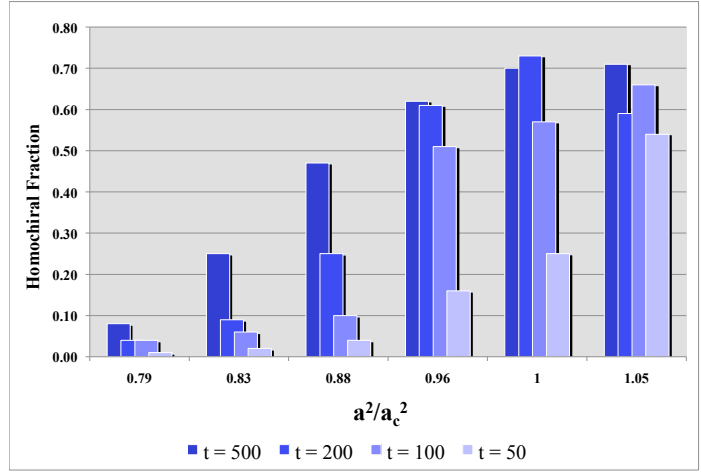}}
\caption{Fraction of $2d$ homochiral worlds after a single environmental event. Colors indicate duration of the event, while the horizontal axis labels its magnitude in units of $a_c^2$, the critical value for chiral symmetry restoration. The statistical sample included $100$ runs and the time scales of the events are in units $t(l_0^{-1})$.} \label{fig:data}
\end{figure}

Fig. \ref{fig:data} summarizes the results of a detailed statistical analysis of $100$ $2d$ runs that led to initial domain coexistence, that is, $d \langle {\cal A} \rangle /dt \approx 0$ (see Fig.  \ref{fig:3event} for $t < 600$) \cite{GTW}. The horizontal axis displays the magnitude of the disturbance in units of $a_c^2$. The vertical axis gives the fraction of homochiral worlds, that is, those that after the disturbance obtain chiral purity. The colors represent the duration of the event. For $a^2 \geq 0.96 a_c^2$, that is, near the critical region, all but the shortest events ($t \leq 50 l_0^{-1} \approx 1.5$ months, for the nominal value of $l_0=\sqrt{2} \times 10^{-5}$s$^{-1}$ mentioned previously) lead to statistically significant chiral biasing. Results in $3d$ are qualitatively very similar \cite{GTW}.

\section{Chirality from Chiral-Selective Reaction Rates \cite{GNW}}

Could prebiotic homochirality be achieved without autocatalysis exclusively through chiral-selective reaction rate parameters without any other explicit mechanism for chiral bias? In a recent work, we investigated this question, focusing on the simplest possible chemical network: no autocatalysis and no enantiomeric cross-inhibition, the trademarks of Frank-like chirobiogenesis models. Instead, we investigated how rare a set of chiral-selective reaction rates in a polymerization model needs to be in order to generate a reasonable amount of chiral bias. We quantified our results adopting a statistical approach: varying both the mean value and the rms dispersion of the relevant reaction rates, we showed that moderate to high levels of chiral excess can be achieved.

Consider a polymerization reaction network where activated levorotatory and dextrorotatory monomers ($L_1^*$ and $D_1^*$, respectively), can chain up to generate longer homochiral or heterochiral molecules. (Here, ``activated'' refers to these monomers being highly reactive due to energy input into the reactor pool.) The model reactions include deactivation of activated monomers,
\begin{align}
L_1^* \stackrel{h_L}\rightarrow L_1  && D_1^* \stackrel{h_D}\rightarrow D_1 
\end{align}
and the polymerization reactions,
\begin{align}
L_1^* + L_i \stackrel{a_L}\rightarrow L_{i+1}  && D_1^* + D_j \stackrel{a_D}\rightarrow D_{j+1} \\
L_1^* + D_j \stackrel{a_L}\rightarrow M_{1j}  && D_1^* + L_i \stackrel{a_D}\rightarrow M_{i1} \\
L_1^* + M_{ij} \stackrel{a_L}\rightarrow M_{i+1j}  && D_1^* + M_{ij} \stackrel{a_D}\rightarrow M_{i j+1}
\end{align}
where $h_{L(D)}$ is the deactivation rate for $L(D)$-monomers and $a_{L(D)}$ is the polymerization rate for adding activated $L(D)$-monomers to a growing chain. Here $L_i$ and $D_j$ denote homochiral polymers of length $i$ and $j$ respectively, and $M_{ij}$ denotes polymers of mixed chirality consisting of $i$ $L$-monomers and $j$ $D$-monomers. There is no autocatalysis (either explicit or through enzymatic activity of homochiral chains) or enantiomeric cross-inhibition present. We also add a source $S$ and a disappearance rate $d$ to the reaction equations so as to model an open system. All reaction rates are scaled by the disappearance rate $d$.

Instead of assigning {\it ad hoc} values for the rates of opposite chirality, we conducted a statistical analysis whereby, for different numerical experiments--that is, different realizations of prebiotic scenarios--the values for $L$ and $D$ reactions rates were picked from a set allowed to randomly fluctuate about a given mean. In other words, each time we solved the coupled nonlinear ordinary differential equations describing the reactions, we picked a set of random, Gaussian-distributed values for the four reactions rates $h_{L(D)}$ and $a_{L(D)}$. (For details, see Ref. \cite{GNW}.) 

We define the enantiomeric excess [$ee(t)$], that is, the net chirality, as

\begin{equation}
\label{ee_evol}
ee(t)=\frac{ [L_1^*(t)+L_1(t)+{\cal L}(t)]-[D_1^*(t)+D_1(t)+{\cal D}(t)]}{[L_1^*(t)+L_1(t)+{\cal L}(t)]+[D_1^*(t)+D_1(t)+{\cal D}(t)]}.
\end{equation}
Note that our definition of the enantiomeric excess implicitly assumes that we are only interested in pure homochiral polymers. Here, we have taken the limit of large $N$ in the polymerization equations and introduced 
\begin{align}
{\cal L} = \sum_{i=2}^\infty L_i,~~~~~ {\cal D} = \sum_{j=2}^\infty D_j,~~~~~{\cal M} =  \sum_{i=1}^\infty \sum_{j=1}^\infty M_{ij}, 
\end{align}
for levorotatory, dextrorotatory and mixed polymers, respectively.

In Figure \ref{ee_evol} we show the behavior of the time evolution of enantiomeric excess for several illustrative examples with $S/d\equiv \sigma = 200$ and different choices for the reaction rates $a_{L(D)}/d\equiv \alpha_{L(D)}$ and $h_{L(D)}/d\equiv \beta_{L(D)}$. On the left, we show results with $\beta_L = \beta_D=100$ fixed, and varying ratios of $\alpha_L/\alpha_D$. Shown are the results for $\alpha_L=5.0$, $\alpha_D=10.0$ ($\alpha_L/\alpha_D = \frac{1}{2}$) in blue, $\alpha_L=5.0$, $\alpha_D=15.0$ ($\alpha_L/\alpha_D = \frac{1}{3}$) in black, and $\alpha_L=5.0$, $\alpha_D=20.0$ ($\alpha_L/\alpha_D = \frac{1}{4}$) in red.  The system quickly reaches a steady state with net chiral excess $ee_{ss}\simeq 0.24$,  $ee_{ss}\simeq 0.38$, and  $ee_{ss}\simeq 0.48$, respectively (corresponding to $24\%$, $38\%$, and $48\%$ enantiomeric excess). On the right, we show results holding  $\alpha_L= \alpha_D=10.0$ fixed, and varying the ratio of $\beta_L/\beta_D$. Shown are the results for $\beta_L = 75$, $\beta_D = 25$ ($\beta_L/\beta_D = 3$) in blue,  $\beta_L = 100$, $\beta_D = 25$ ($\beta_L/\beta_D = 4$) in black, and $\beta_L = 125$, $\beta_D = 25$ ($\beta_L/\beta_D = 5$) in red. In this case the net chiral excess at steady-state is $ee_{ss}\simeq 0.38$,  $ee_{ss}\simeq 0.47$, and  $ee_{ss}\simeq 0.58$, respectively (corresponding to $38\%$, $47\%$, and $58\%$ enantiomeric excess). These test runs show that for differing left and right reaction rates a substantial amount of chiral excess may be reached at steady-state. Although for these examples the differences were fairly large, we note that if the chiral-selective changes in rates appear in exponential factors (as in the example above for temperature dependence), small changes in the parameters may generate fairly large changes in the resulting reaction rates. 

\begin{figure}[htbp]
\includegraphics[width=0.49\linewidth]{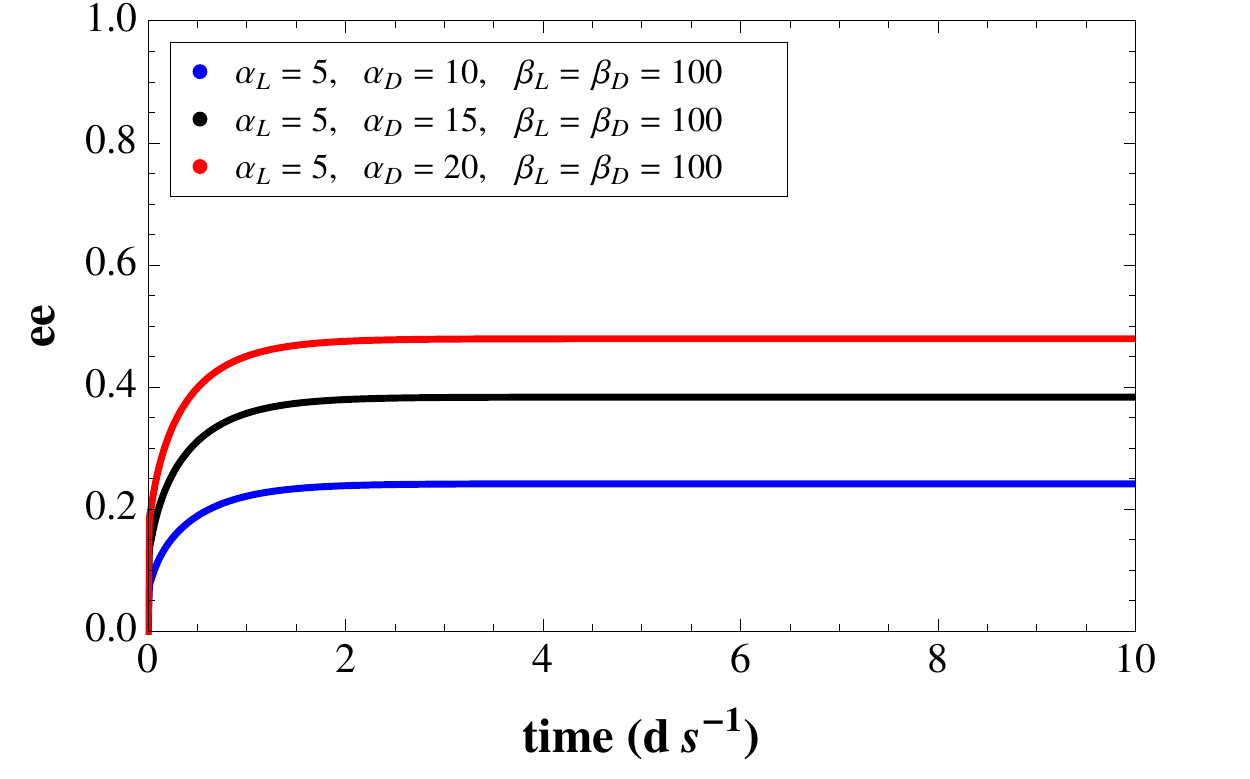}
\includegraphics[width=0.49\linewidth]{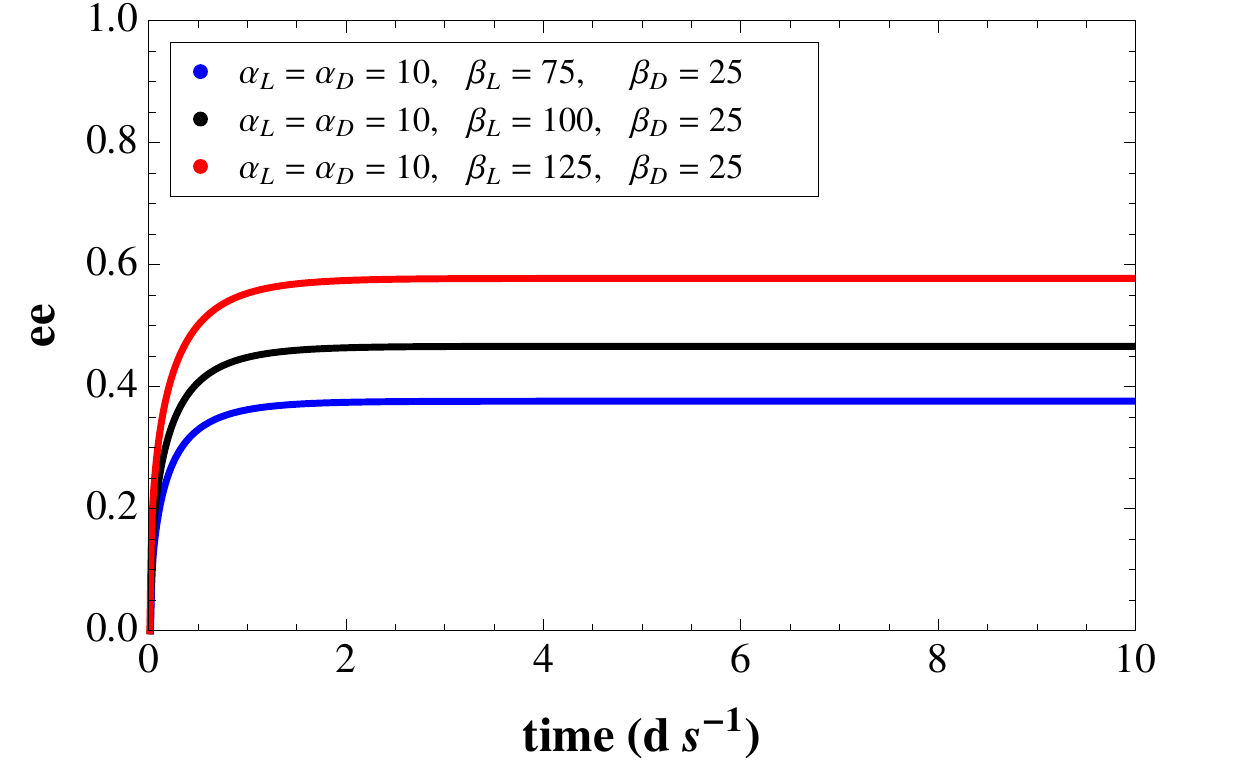} 
\caption{Time evolution of enantiomeric excess for illustrative examples of reaction rates with $\sigma = 200$. {\bf Left:} $\beta_L = \beta_D=100$, with varying ratios of $\alpha_L/\alpha_D$. Shown are $\alpha_L=5.0$, $\alpha_D=10.0$ ($\alpha_L/\alpha_D = \frac{1}{2}$) in blue, $\alpha_L=5.0$, $\alpha_D=15.0$ ($\alpha_L/\alpha_D = \frac{1}{3}$) in black, and $\alpha_L=5.0$, $\alpha_D=20.0$ ($\alpha_L/\alpha_D = \frac{1}{4}$) in red. {\bf Right:} $\alpha_L= \alpha_D=10.0$, with varying ratios of $\beta_D/\beta_L$. Shown are  $\beta_L = 75$, $\beta_D = 25$ ($\beta_L/\beta_D = 3$) in blue,  $\beta_L = 100$, $\beta_D = 25$ ($\beta_L/\beta_D = 4$) in black, and $\beta_L = 125$, $\beta_D = 25$ ($\beta_L/\beta_D = 5$) in red}
\label{ee_evol}
\end{figure}

We have performed a detailed statistical study of the reaction equations with fluctuating values for the $L$-- and $D$-- reaction rates $\alpha_L$, $\alpha_D$, $\beta_L$ and $\beta_D$. This involves solving the coupled system of ordinary differential equations $N$ times, each with values for the four reaction rates given by $\alpha_L= \bar \alpha+\delta_L$, $\alpha_D= \bar \alpha +\delta_D$, $\beta_L= \bar \beta+\xi_L $, and $\beta = \bar \beta_D+\xi_D$, where the bars denote the mean values and $\delta_{L(D)}$ and $\xi_{L(D)}$ are Gaussian-distributed random numbers, within a fixed rms width set by $\alpha_0$ and $\beta_0$. 

\begin{figure}[htbp]
\centerline{\includegraphics[scale=0.75]{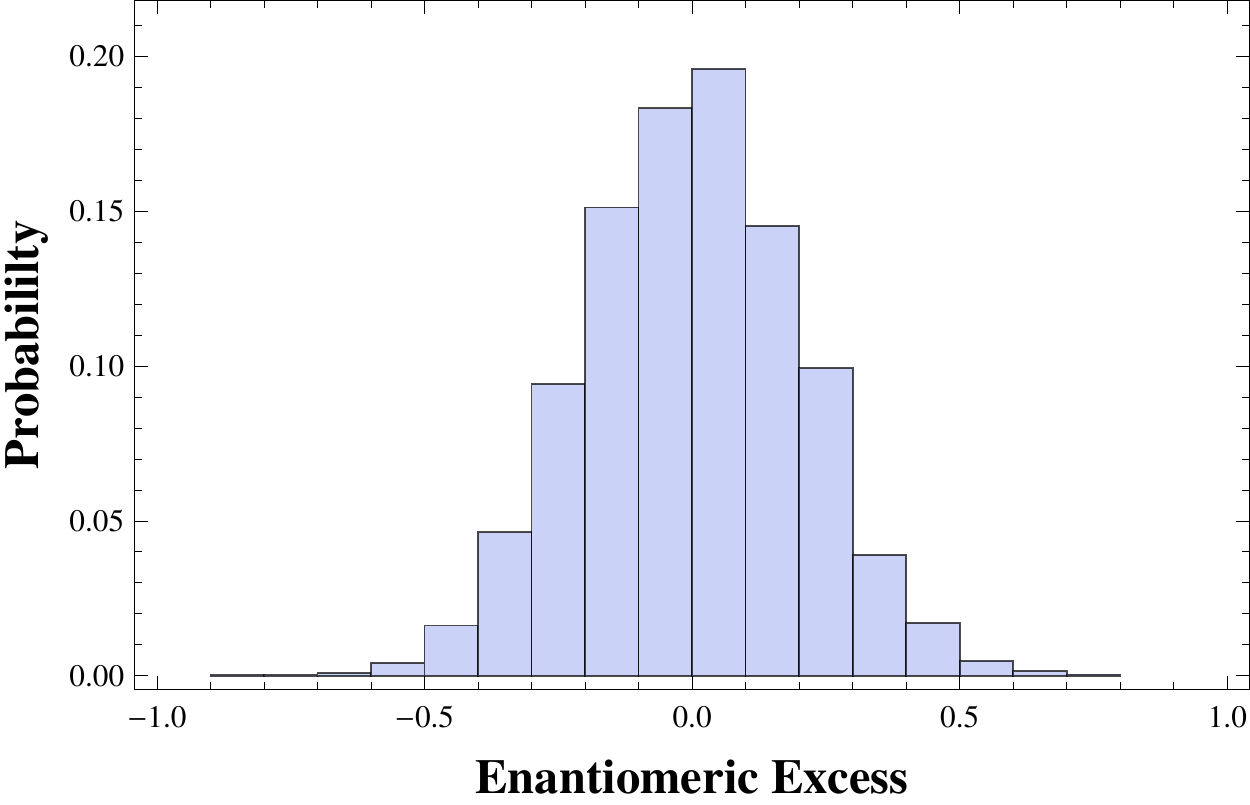}}
\caption{Distribution of ensemble $ee$ values for $N = 5000$ with $\bar{\alpha} = 20$, $\bar{\beta} = 120$, $\alpha_0 = \beta_0 = 0.25$, and $\sigma = 200$ .}
\label{probability_dist}
\end{figure}

Our results indicate that the net chirality is overall more sensitive to the amplitude of the chiral-selective variations about the mean values of the reaction rates than to the values of the reaction rates themselves. 
In Figure \ref{probability_dist}, we explore the spread in the distribution of the enantiomeric excess in experimental systems for an ensemble with $\bar{\alpha} = 20$, $\bar{\beta} = 120$ and $\alpha_0 = \beta_0 = 0.25$. Outliers in the distribution have very high enantiomeric excesses up to as much as $80 - 90 \%$, although most systems fall within a range of enantiomeric excesses with $|ee| < 0.25$, where the mean of the distribution lies. 

\section{Conclusions}

During the past decades, several biasing mechanisms have been proposed to explain life's remarkable homochirality. Parity violation in weak neutral currents \cite{Yamagata}, if effective, would provide a universal bias: all amino acids found in the universe should be levorotatory. (Although one would still need to explain why sugars are dextrorotatory.) In contrast, circularly-polarized UV light \cite{Lucas}, if produced in active star-forming regions, would act within a stellar system or, at most, within neighboring stellar systems {\it without} any uniform bias: in different star-forming regions across the galaxy, stellar systems should have stereochemistry with uncorrelated chirality. One of us has recently argued that--if relegated to diffusive processes--both mechanisms would probably be ineffective within the time-scales relevant for life's emergence on Earth \cite{G}. In any case, the point we are making here is stronger: punctuated chirality would render {\it any} biasing mechanism ineffective: environmental events have the potential to restore chiral symmetry and thus to wash out previous values of chirality locally and, for events of great violence, globally. The same conclusion holds for the very different mechanism toward homochirality  described in the last section, where no autocatalysis or enantiomeric excess is needed to amplify an initial bias; instead, chiral-selective reaction rates alone can lead to substantial chiral excess. We noted that this chiral selectivity may also be prompted by environmental effects \cite{GNW}. Both parity violation and circularly-polarized UV light predict that within the same stellar system, each planetary platform would have the same chiral bias. In contrast, environmentally-induced chiral selectivity is a local effect, that depends on the particular history of the planetary platform harboring life. We thus propose that it is possible to distinguish between the three mechanisms through future space missions aimed at studying stereochemistry \cite{Lunine}. If chiral bias, as life on Earth, goes through periods of stasis (chemical steady state) punctuated by violent upheavals and symmetry restoration, we predict that there would be no chiral correlations even within the same stellar system: the same amino acid found, say, in Titan would not necessarily display the same chirality if found on Earth or Mars. Of course, only a large enough statistical sample would resolve the issue.

Biological precursors certainly interacted with the primordial environment and may have had their chirality reset multiple times before homochiral life first evolved. In this case, separate domains of molecular assemblies with randomly set chirality may have reacted in different ways to environmental disturbances. A final, Earth-wide homochiral prebiotic chemistry would have been the result of multiple interactions between neighboring chiral domains \cite{BM,GW,G} in an abiotic process that mimics natural selection.

\section*{Acknowledgments}
MG is supported in part by a National Science Foundation grant PHY-1068027. SIW gratefully acknowledges support from the NASA Astrobiology Institute through the NASA Postdoctoral Fellowship Program.

\end{document}